\documentstyle[prl,aps,twocolumn,psfig]{revtex}
\begin{document}

\title
{Superconducting single-mode contact as a microwave-activated
quantum interferometer}

\date{\today}
\author{L. Y. Gorelik$^{1,3}$, N. I. Lundin$^{1}$, V. S. Shumeiko$^{2,3}$, 
R. I. Shekhter$^{1}$, and M. Jonson$^{1}$}
\address{
 $^{1}$Department of Applied Physics and $^{2}$Department of Microelectronics 
and Nanoscience, 
Chalmers University of Technology and G\"oteborg University, 
S-412 96 G\"oteborg, Sweden; 
 $^{3}$B. I.  Verkin Institute for Low Temperature Physics \& Engineering, 
National Academy of Science of Ukraine, 4 Lenin Ave., 310164 Kharkov, Ukraine
\\
\vspace{5mm}
{\parbox{14cm} {\small
The dynamics of a superconducting quantum point contact biased at subgap 
voltages is shown to be strongly affected by a microwave electromagnetic field. 
Interference among a sequence of temporally localized, microwave-induced 
Landau-Zener transitions between current carrying Andreev levels results 
in energy absorption and in an increase of the subgap current by several 
orders of magnitude. The contact is an interferometer in the sense that 
the current is an oscillatory function of the inverse bias voltage. 
Possible applications to Andreev-level spectroscopy and microwave 
detection are discussed.}}
}
\maketitle


\vspace{0.3cm}
\narrowtext

The classic double-slit interference experiment, where two spatially separated
trajectories combine to form an interference 
pattern, clearly demonstrates the wavelike nature of electron propagation.
For a 0-dimensional system, with no spatial structure, a completely analogous
interference phenomenon may occur between two distinct trajectories in the 
{\em temporal} evolution of a quantum system. Such trajectories may appear
in the presence of temporally localized non-adiabatic perturbations (rather 
than spatially localized slits in a screen) which scatter the system from
one adiabatically evolving state to another. In this Letter we show that this 
type of interference phenomenon significantly controls the microscopic dynamics 
of a voltage-biased superconducting quantum point contact (QPC) subject to 
microwave irradiation. 

It is well known that the Josephson current in a QPC is carried
by Andreev bound states localized within the contact area. The corresponding
energy levels --- the Andreev levels --- lie in the
energy gap of the superconductor and their positions depend on the change $\phi$
in the phase of the condensate across the junction. Hence, the Andreev levels
will move adiabatically with time within the gap if the contact is biased by a 
voltage $V$ much smaller than the gap energy, $\Delta$. With any (normal) 
electron scattering present in the contact the Andreev levels will, however,
never cross the Fermi level; instead they will oscillate periodically with 
$\phi$ so that on the average no energy is transfered 
to the QPC and a purely ac current will flow through the contact 
(ac Josephson effect).

Microwave radiation of large frequency $\omega\sim\Delta$ represents 
a non-adiabatic perturbation of the QPC system. However, if the amplitude 
of the electromagnetic field is sufficiently small, the field will not affect
the adiabatic dynamics of the system much unless the condition for resonant
optical interlevel transitions is fulfilled. Such resonances will only occur 
at certain moments determined by the time evolution of 
the Andreev level spacing. 
They provide a mechanism for the energy transfer to the system to be nonzero
when averaged over time and for a finite dc current through the junction. 
The rate of energy transfer is in an essential way
determined by the interference between different scattering events \cite{us},
which will also lead to oscillatory features in the current-voltage 
characteristics of the QPC.

Although an example of the more general problem of energy level spectroscopy,
the spectroscopy of Andreev levels has an important specific aspect. Due to
their ability to carry electric current, detection
of optical transitions between Andreev levels is possible by means of transport
measurements. The appearance of a subgap current 
under resonant radiation can furthermore be used as a sensitive microwave 
detector.

\begin{figure}
\centerline{\psfig{figure=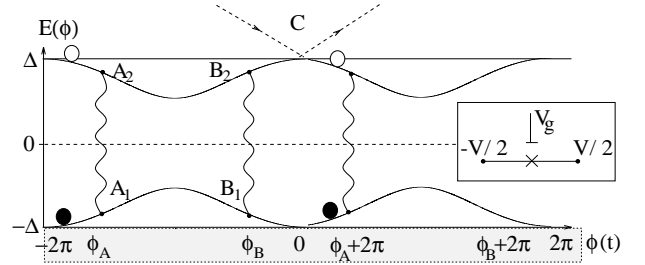,width=8 cm}}
\vspace{3mm}
 \caption{ \label{fig:junction}
Time evolution of Andreev levels (full lines) in the energy gap of a 
gated voltage-biased, single-mode superconducting QPC (see inset). 
A weak microwave field induces 
resonant transitions (wavy lines) between the levels at points $A$ and $B$ 
and the level above the Fermi energy becomes partly occupied  to an extent 
determined by interference between the two transition amplitudes. 
Non-adiabatic interactions release the energy of quasiparticles in the 
(partly) occupied Andreev level into the continuum 
at point $C$, where the Andreev states and the continuum merge into each other (represented by dashed arrows, see text) and the initial conditions for the 
Andreev level populations are reset (filled and empty circles). 
}
\end{figure}

For an unbiased QPC, the Andreev spectrum of each transport mode has the form
$E^\pm(\phi)=\pm E(\phi)=\pm\Delta[1-D\sin^2(\phi/2)]^{1/2}$, where
$D$ is the transparency of the mode and the energy is measured from the 
Fermi energy \cite{Fur,Bee}. With a small bias voltage applied, the levels 
move along the adiabatic trajectories $E^\pm(t)=\pm E(\phi_0+2eVt/\hbar)$
in energy-time space, as shown in Fig.~1. 
When the criterium $\hbar\dot\phi=2eV\ll 2E^2(t)/\Delta$ for adiabaticity is obeyed,
the rate of interlevel transitions is exponentially small keeping the
level populations constant in time~\cite{Ave1,Lv1}. 
The presence of a weak electromagnetic field [on the scale of $ E(t)$]
does not affect the adiabatic level trajectories except for short times
close to the resonances at $t=t_{A,B}$, when $E(t_{A,B})=\hbar\omega/2$.
Here the dynamics of the system is strongly
non-adiabatic with a resonant coupling which effectively mixes
the adiabatic levels. This is an analog of the well known Landau-Zener
transition, which describes interlevel scattering as a resonance point 
is passed. In our case these transitions give rise to a splitting of 
the quasiparticle trajectory at the points $A, B$ into two paths; $A_1A_2B_2$ 
and $A_1B_1B_2$, forming a loop in $(E,t)$ space.

The resonant scattering opens a channel for energy absorption by the system;
a populated upper 
level when approaching the edge of the 
energy gap (at point $C$ in Fig.~1) creates real excitations in the continuum 
spectrum, which carry away
the accumulated energy from the contact. As a result, the net rate of 
energy transfer to the system is finite; it consists of energy absorbed 
both from the electromagnetic field and from the voltage source. 
The confluence of the two adiabatic trajectories at $B_2$ (see Fig.~1) 
gives rise to a strong interference pattern in the probability for 
real excitations at the band edge, point $C$. The interference effect is 
controlled by the difference of the phases acquired by the system during 
propagation along the paths $A_1A_2B_2$ and $A_1B_1B_2$.

For a quantitative discussion we consider a one-mode
superconducting point contact~\cite{many-mode} with arbitrary energy-independent
transparency $D$ for normal electrons, $0<D<1$. The contact is biased at
a small applied voltage $eV\ll\Delta$, and a high frequency electromagnetic
field is applied to the gate situated near the contact,
see inset in Fig.~1.
We will describe the evolution of the Andreev states with the time-dependent
Bogoliubov-de Gennes equation 
\cite{book} for the quasiclassical envelopes $u_\pm(x,t)$ of
the two-component wave function
$\Psi(x,t)=u_+(x,t)e^{i k_Fx} + u_-(x,t)e^{-i k_Fx}$,
\begin{equation}
i\hbar \partial{\bf u}/\partial t =[{\bf H}_0 + \sigma_z V_g(x,t)] {\bf u} \, .
\label{BdG}
\end{equation}
In this equation ${\bf u}=(u_+, u_-)$ is a four-component vector,
${\bf H}_0$ is the Hamiltonian of the electrons in the electrodes of the point
contact,
\begin{eqnarray}
\label{H0}
{\bf H}_0&=&-i\hbar v_F\sigma_z\tau_z\partial/\partial x + \\
&&\Delta[\cos(\phi(t)/2)\sigma_x + i\sin(\phi(t)/2)\mbox {sgn} x\sigma_y],
\nonumber
\end{eqnarray}
where $\sigma_i$ and $\tau_i$ denote Pauli matrices in
electron-hole space and in $u_\pm$ space respectively.
The function ${\bf u}$, which is smooth on the scale of the Fermi wavelength, 
has a discontinuity at the contact which is determined by the transfer
matrix of the QPC in the normal state and is described by the following 
boundary condition~\cite{Sh}:
%
${\bf u}(+0)=(1/\sqrt D)(1-\sqrt R\tau_y){\bf u}(-0), \;\; R=1-D.$
%

The gate potential $V_g(x,t)=V_\omega(x)\cos \omega t$
in Eq.~(\ref{BdG}) oscillates rapidly in time and
the amplitude 
is assumed to be
small compared to the Andreev level spacing, $V_\omega\ll E(t)$. Under
this condition, the system experiences an adiabatic evolution at all times
except close to the resonances (points $A$ and $B$ in Fig.~1) 
The duration $\delta t$ of these resonances is short in the limit 
$eV\ll\Delta$ compared to the period of Josephson oscillations
$T_V=\pi\hbar/eV$. Indeed, the resonant transition occurs if the deviation 
of the interlevel spacing from the resonance value
$E(t)-\hbar\omega/2=\dot E(t_{A,B})\delta t$ does not exceed the
quantum mechanical resolution of the energy levels $\hbar/\delta t$. From 
this we can estimate 
$\delta t$ to 
$\delta t\sim [\hbar/\dot E(t_{A,B})]^{1/2}\ll\hbar/eV$. Hence we may
consider the non-adiabatic dynamics as temporally localized scattering events.
By introducing a linear combination 
%
${\bf u}(t)=b^+ (t)e^{i\omega t/2}{\bf u}^+ + b^-(t) e^{-i\omega t/2}{\bf u}^- $
%
of the eigenstates
${\bf u}^\pm$ corresponding to the adiabatic Andreev levels $E^\pm$, we can 
describe the system's evolution through a resonance by letting
a scattering matrix $\hat S$ connect the
coefficients $b^\pm$ before and after the 
splitting points $A$ and $B$. A standard analyses of the Landau-Zener
interlevel transitions (see e.g. \cite{smslt}) gives the scattering matrix
elements at the point $A$ $(S_A)_{++}=(S_A)_{--}=\tau$,
$(S_A)_{+-}=-(S_A)^\ast_{-+}=i\rho $, where
$|\rho|^2 =1-|\tau|^2=1-e^{-\gamma}$ is the probability of the 
Landau-Zener interlevel transition.
Here $\gamma=\pi |V_{+-}|^2/(dE/dt)$
where $V_{+-}$ is the matrix element for the interlevel transitions.
At the splitting point $B$ the
scattering matrix reads $\hat S_B =\hat S^T_A$. The matrix element $V_{+-}$
was calculated for the case of a double barrier QPC structure
in Ref.~\cite{smslt}. For a single barrier
junction an analogous calculation gives us, 
$V_{+-}=\alpha(L/\xi_0)\sqrt{DR}V_\omega\sin(\phi/2)$, where
$L$ is the length of the normal region, $\xi_0$ is the coherence length, and 
the constant $\alpha\sim 1$ is determined by the position of the barrier.
We note, that this matrix element 
is proportional to the {\em reflectivity} of the junction and therefore 
the effect under consideration does not exist in perfectly transparent QPC 
($D=1$), cf. Refs. \cite{Ave1,Zai}.

By introducing the matrix $\hat\Phi_{j,i}=\exp{(i\sigma_z\Phi(i,j))}$, 
\begin{equation}
\Phi(i,j)= {1\over 2eV}\int_{\phi_i}^{\phi_j}
d\phi \left( E(\phi)-{\hbar\omega\over 2}\right),
\label{phi}
\end{equation}
which describes the ``ballistic" 
dynamics of the system between the Landau-Zener
scattering events, we connect the coefficients $b^\pm$ at the end of the
period of the Josephson oscillation, $\phi=0$, with the coefficients 
$b^\pm_0$ at the beginning of the period, $\phi=-2\pi$,
\begin{equation}
{b^+\choose b^-}=\hat\Phi_{0,B}\hat S_B\hat\Phi_{B,A}\hat S_A\hat\Phi_{A,-2\pi}
{b^+_0\choose b^-_0}.
\label{bb0}
\end{equation}
%
The time-averaged current through the junction can be directly expressed 
through these coefficients.

The quasiclassical equation for the total time dependent current at the
junction ($x=0$) reads 
$I(t)=v_F\langle{\bf u}\sigma_z\tau_z{\bf u}\rangle$, 
where $\langle..\rangle$ denotes a scalar product in 4-dimensional space.
From Eqs.~(\ref{BdG},\ref{H0}) it follows that,
\begin{equation}
I(t)=\frac{2e}{\hbar}\left({d\phi\over dt}\right)^{-1}\int^\infty_{-\infty}
dx\left[ i\hbar \frac{d\langle{\bf u}\dot {\bf u}\rangle}{dt}-
\dot V_g\langle{\bf u}\sigma_z{\bf u}\rangle
\right].
\label{j}
\end{equation}
In the static limit, $\dot V_g=0$ and $\dot\phi\rightarrow 0$, it equals 
the usual equation $I=(2e/\hbar)\,(dE^\pm/d \phi)$
for the Andreev level current. In the general nonstationary case ${\bf u}$ 
is a linear combination of ${\bf u}^\pm$ and we
%
%
calculate the current averaged over the period $T_V$.
Using the normalization condition $|b^+|^2+|b^-|^2=1$
and omitting small contributions from rapidly oscillating terms, 
we obtain
\begin{equation}
I_{dc}={2e\over \pi\hbar}\left(\Delta-{\hbar\omega\over 2}\right)\left[
|b^+|^2-|b^+_0|^2\right].
\label{Idc}
\end{equation}
The direct current through the contact can be viewed as resulting from
photon-assisted pair tunneling or equivalently as being due to the
distortion of the ac pair current due to the induced interlevel
transitions.
The magnitude of the current is such that the energy absorbed from
the voltage source, $V\cdot I_{dc}$, together with the energy absorbed
from the hf field corresponds to the energy necessary for creating a
real continuum-state excitation.

Let us now discuss the boundary condition at $\phi=2\pi n$ ($n$
is an integer). In the vicinity of these points, the Andreev levels
approach the continuum and the adiabatic approximation is unsatisfactory,
even at small applied voltages and weak electromagnetic fields.
The duration $\delta t$ of the non-adiabatic interaction between the Andreev
level and the continuum states can be estimated 
using the same argument as for the microwave-induced Landau-Zener scattering. 
One finds that $\delta t \sim \hbar/(\Delta e^2V^2)^{1/3}$.
To derive the boundary condition, for example, at point $C$ in Fig.~1, one
needs to calculate the transition amplitude connecting the states 
${\bf u^+}(t_1)$ at time 
$t_1\ll t_C-\delta t$ and ${\bf u^+}(t_2)$
at time $t_2\gg t_C+\delta t$: 
$\langle {\bf u^+}(t_2)\;{\bf U}(t_2,t_1){\bf u^+}(t_1)\rangle$. Here
${\bf U}(t_2,t_1)$ is the exact propagator corresponding to the Hamiltonian 
in Eq. (\ref{BdG}). It follows from symmetry arguments that this
amplitude is exactly zero.
Both the Hamiltonian (\ref{H0}) and the boundary condition for ${\bf u}$ at
$x=0$
are invariant under the simultaneous charge- and parity inversion
described by the unitary  operator ${\bf\Lambda}\equiv\hat P\sigma_x\tau_z$,
where $\hat P$ is the parity operator in $x$-space. This implies that 
at any time any non-degenerate eigenstate of the 
Hamiltonian is an eigenstate of 
the symmetry operator ${\bf\Lambda}$ with eigenvalue $+1$ or $-1$ and that
this property persists during the time evolution of the state. In particular,
the static Andreev state obeys the equation 
${\bf \Lambda}{\bf u^+}(\phi)=\Lambda{\bf u^+}(\phi)$
at any $\phi$. It follows from Eq.~(\ref{H0}) that 
${\bf u^+}(-\phi)=\sigma_x\tau_y {\bf u^+}(\phi)$ so
$\Lambda(-\phi)=-\Lambda(\phi)$. The immediate consequence of this property 
is that the static Andreev states correspond to {\em different}
eigenvalues $\Lambda$ at $\phi<0$ and at $\phi>0$, and therefore they are 
orthogonal \cite{period}, even though they belong to the {\em same}
level.
Hence the state evolving from the adiabatic state ${\bf u^+}(t_1)$ is
orthogonal to the adiabatic state ${\bf u^+}(t_2)$. 
As a result the probability
for an adiabatic Andreev state to be ``scattered" into a
localized state after passing the non-adiabatic region is identically zero.
In reality,
the Andreev state as it approaches the continuum band edge 
decays into the states of the continuum.  Such a decay corresponds to a
delocalization in real space and is the mechanism for
transferring energy to the reservoir \cite{LV}.

The orthogonality property shown above guarantees that the coherent
evolution of our system persists during only one period of the Josephson
oscillation and that the equilibrium population of the Andreev levels is
reset at each point $\phi=2\pi n$ \cite{Ave}. This imposes the boundary
conditions $b^+(2\pi n+0)=0,\;b^-(2\pi n+0)=1$ in the beginning of each
period. Combining this boundary condition with Eqs. (\ref{Idc}) and
(\ref{bb0}), we finally get
\begin{equation}
I_{dc}={8e\over \pi\hbar}\left(\Delta-{\hbar\omega\over 2}\right)e^{-\gamma}(1-e^{-\gamma})\sin^2(\Phi(A,B)+\theta)
\label{I}
\end{equation}
where $\theta$ is the phase of the probability
amplitude for the Landau-Zener
transition, which weakly depends on $V$.
\begin{figure}
\vspace{-.8cm}
\centerline{\psfig{figure=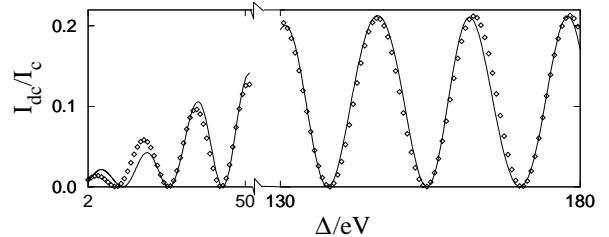,width=8 cm}}
 \caption{ \label{fig:compare}
Current vs. inverse voltage from Eq.~(7) for a biased superconducting QPC 
irradiated with microwaves of frequency $\omega=1.52\Delta/\hbar$ and 
amplitude corresponding to a matrix element $|V_\pm|=0.024\Delta$ 
for interlevel transitions. 
Note the cut in the inverse voltage scale.
Results of the scattering approach, 
Eq.~(5) [$\diamond$], are close to those obtained by numerically solving 
the BdG-equation (1) with the radiation field treated in the resonance 
approximation (solid line).
The close fit means that the scattering picture can be used to reconstruct
the Andreev levels from the period of the current oscillations (Andreev level 
spectroscopy, see text). 
}
\end{figure}

Equation (\ref{I}) is the basis for presenting the biased QPC 
as a quantum interferometer. There is a clear analogy between the 
QPC interferometer and a standard SQUID in that they both rely on the presence 
of trajectories that form a closed loop. In a SQUID, which is used to measure
magnetic fields, the loop is determined by the device geometry; in the QPC the 
voltage (analog of the magnetic field) is well defined while the ``geometry"
of the loop in $(E,t)$ space can be measured. This loop is determined by
the Andreev-level trajectories in $(E,t)$ space and is controlled by the
frequency of the external field. This gives us an immediate possibility to
reconstruct the phase dependence of the Andreev levels from the frequency
dependence of the period $\Pi$ of oscillations of the current versus inverse 
voltage, see Fig.~2. Indeed, it follows from Eq.~(\ref{phi}), that
\begin{equation}
\phi(E)=\pi\pm {4\pi e\over\hbar}
\left.{d\Pi^{-1}\over d\omega}\right|_{\omega=2E}.
\end{equation}

In order to be able to do interferometry it is necessary to keep
phase coherence during one period of the Josephson oscillation. There are
three dephasing mechanisms that impose limitations in practice: (i)
deviations from an ideal voltage bias, (ii) microscopic interactions, and
(iii) radiation induced transitions to continuum states. The main source of
fluctuations of the applied voltage is the ac Josephson effect.
According to the RSJ model, a fixed voltage across the
junction can only be maintained if the ratio between the intrinsic 
resistance $R_i$ of the voltage source and the normal junction
resistance $R_N$ is small. If
$R_i/R_N\ll 1$ the amplitude of the voltage fluctuations $\delta V$ is
estimated as $\delta V\sim (R_i/R_N)V$. Effects of voltage fluctuations
on the accumulated phase $\Phi(A,B)$ can be neglected if 
$\delta \Phi(A,B)=\Phi'(A,B)\delta V\ll 2\pi$, i.e. if $eV>(R_i/R_N)\Delta$,
which corresponds to a lower limit for the bias voltage voltage \cite{C}.

The dephasing time due to microscopic interactions is comparable to the
corresponding relaxation time \cite{Gefen}. This mechanism of dephasing 
can be neglected as soon as the relaxation time exceeds the Josephson 
oscillation period, $\tau_i\gg \hbar/eV$. 
Taking electron-phonon interaction as the leading mechanism of inelastic
relaxation, we estimate $\tau_i$ to be of the order of the electron-phonon
 mean free time at the critical temperature, $\tau_{ph}(T_c)$, since the
large deviations from equilibrium in our case occur in the energy interval
$E<\Delta$. This gives \cite{Kaplan} another limitation on how small 
the applied voltage van be, $eV > 10^{-2}\Delta$.

The third mechanism of dephasing becomes important when the Andreev 
levels are closer 
than $\hbar \omega$ to the continuum band edge. One can estimate the 
corresponding relaxation time as $\tau_\omega \sim \hbar \Delta/V^2_\omega$. 
For small radiation amplitudes $\tau_\omega$ exceeds $T_V$, while for 
optimal amplitudes they are about equal. The effect of the level-continuum 
transitions on the interference oscillations depends on the frequency.
If $\hbar \omega < 2\Delta /3$ the ``loop region" [$E(t) < \hbar \omega /2$]
is optically disconnected from the continuum, and 
transitions cannot destroy interference. Possible level-continuum transitions
at times outside the loop will only decrease the amplitude of the effect by a
factor $\exp(-\alpha V_\omega^2/eV\Delta)$, where $\alpha < 1$ is the relative 
fraction of the period $T_V$ during which transitions to the
continuum are possible. Accordingly, this factor
is of the order of unity for the voltages that correspond to the
maximum amplitude of oscillations. If $\hbar \omega > 2\Delta /3$,
the interference is impeded by the optical transitions into the
continuum, and the current oscillations decrease. Still,
a nonzero average current through the junction will persist.

The interference effect presented here can also be applied for detecting
weak electromagnetic signals up to the gap frequency. Due to the resonant
character of the phenomenon, the current response is proportional to the
ratio between the amplitude of the applied field and the applied voltage,
$I\sim|V_\pm|^2/\Delta eV$. At the same time,
for common SIS detectors a non-resonant current response 
is proportional to the ratio between the amplitude and the frequency
of the applied radiation \cite{Tucker}), $I\sim |V_{\pm}/\omega|^2$, i.e. 
it depends entirely on the parameters of the external signal
and cannot be improved.

In conclusion, we have shown that irradiation of a voltage biased
superconducting QPC at frequencies $\omega\sim\Delta$ can 
remove the suppression of subgap dc transport through Andreev levels. 
Due to the resonant nature of the photon-induced interlevel scattering the 
phenomenon can be used for sensitive microwave detection. Quantum interference 
among the resonant scattering events can be used
for microwave spectroscopy of the Andreev levels.

We would like to thank Dag Winkler for productive discussions. 
Support from the Swedish KVA, SSF, Materials consortia 9 and 
11, and NFR is greatfully acknowledged.


\begin{thebibliography}{999}
\bibitem{many-mode}
Our discussion is also valid for a single dominant 
mode in a multi-mode junction.
\bibitem{period}
In fact, Andreev level wave
functions are $4\pi$-periodic with respect to $\phi$ while the energy 
of Andreev levels and the current are $2\pi$-periodic (cf. explicit
equations in Ref. \cite{Sh}).
\bibitem{LV}
This conclusion is in agreement with the analysis of multiple Andreev
reflections in Ref.~\cite{Lv1}.
\bibitem{Ave}
This result is obvious for a pure SNS contact ($D=1$) since the two
Andreev states involved in the transition at the band edge have momenta
$k_F$ and $-k_F$ pointing in opposite directions. See also~\cite{Ave1}.
\bibitem{C}
Shunting by the junction capacitance $C$ is small at relevant frequences
($\sim\!1/\delta t\!\ll\!\Delta$) if
$C\!\leq1\!/R_N\!\Delta\!\sim\!10^{-15}\!F$.
\end{thebibliography}
\end{document}